\documentclass[aps,rpl,preprint,groupedaddress]{revtex4}
\begin{document}

\title{Quantum teleportation using three-particle entanglement}
\author{Ye Yeo}

\affiliation{Centre for Mathematical Sciences, Wilberforce Road, Cambridge CB3 0WB, United Kingdom}

\begin{abstract}
We investigate the teleportation of a quantum state using a three-particle entangled W state.  We compare and contrast our results with those in Ref.\cite{Karlsson} where a three-particle entangled GHZ state was used.  The effects of white noise on the average teleportation fidelities are also studied.
\end{abstract}

\maketitle

The linearity of quantum mechanics allows building of superposition states of composite system that cannot be written as products of states of each subsystem.  Such states are called entangled.  States which are not entangled are referred to as separable states.  An entangled composite system gives rise to nonlocal correlation between its subsystems that does not exist classically.  This nonlocal property enables the uses of local quantum operations and classical communication to transmit information with advantages no classical communication protocol can offer.  The understanding of entanglement is thus at the very heart of quantum information theory \cite{Nielsen}.  In recent years, three-particle entangled states have been investigated by a number of authors \cite{Vidal, Dur, Acin, Andrianov, Rajagopal}.  They have also been shown to have advantages over the two-particle Bell states in their application to dense coding \cite{Wiesner, Hao}, teleportation \cite{Bennett, Karlsson}, and cloning \cite{Buzek, Ekert}.

In Refs. \cite{Vidal, Dur}, Dur {\it et al.} pointed out that the three-particle entangled GHZ state \cite{Greenberger}, while maximally entangled, is not robust in that if one of the three particles is traced out, the remaining two-particle system is not entangled as measured by several criteria.  On the other hand, another three-particle entangled W state \cite{Zeilinger}, which is inequivalent to the GHZ state under stochastic local operations and classical communication, is robust in that it remains entangled even after any one of the three particles is traced out.  More recently, Sen {\it et al.} \cite{Dagomir} showed that $N$-particle entangled W states, for $N > 10$, lead to more ``robust'' (against white noise admixture) violations of local realism, than $N$-particle entangled GHZ states.  The GHZ and W states thus exhibit very different properties when subjected to physical processes like state loss, or white noise.  In Ref.\cite{Karlsson}, A. Karlsson and M. Bourennane demonstrated teleportation of a quantum state using three-particle GHZ state.  In this paper, we review the Karlsson-Bourennane quantum teleportation scheme \cite{Karlsson}, recasting it in the language of density operators and quantum operations.  We then study the consequences of replacing the three-particle GHZ state in their scheme with a three-particle W state.  Finally, we investigate how the presence of white noise affects the teleportation capability of the three-particle GHZ and W states.  We conclude with some remarks on future research.

We begin with a review of the quantum teleportation protocol $P_0$ of A. Karlsson and M. Bourennane \cite{Karlsson}.  It involves a sender, Alice, an accomplice, Cindy, and a receiver, Bob.  Alice is in possession of two two-level quantum systems, the input system 1, and another system 2 maximally entangled with both a third two-level system 3 in Cindy's possession, and a fourth two-level target system 4 in Bob's possession (i.e. a three-particle entangled GHZ state).  Initially the composite system 1234 is prepared in a state with density operator
$$
\sigma^{total}_{1234} = \pi_1 \otimes \chi^{GHZ}_{234}
$$
where
\begin{equation}
\pi_1 = |\psi\rangle_1\langle\psi|,\
|\psi\rangle_1
= \cos\frac{\theta}{2}|0\rangle_1 + e^{i\phi}\sin\frac{\theta}{2}|1\rangle_1,
\end{equation}
$0 \leq \theta \leq \pi,\ 0 \leq \phi \leq 2\pi$ are the polar and azimuthal angles respectively, and
\begin{equation}
\chi^{GHZ}_{234} = |GHZ\rangle_{234}\langle GHZ|,\
|GHZ\rangle_{234} = \frac{1}{\sqrt{2}}(|000\rangle_{234} + |111\rangle_{234}).
\end{equation}
Here, we use $|0\rangle$ and $|1\rangle$ to denote an orthonormal set of basis states for each two-level system.  To teleport the input state $\pi_1$ to Bob's target system 4, Alice performs a joint Bell basis measurement on systems 1 and 2, described by operators $\Pi^j_{12} \otimes I_{34}$, $I_{34}$ is the identity operator on the composite subsystem 34, $j$ labels the outcome of the measurement,
\begin{equation}
\Pi^1_{12} = |\Phi^+\rangle_{12}\langle\Phi^+|,\
\Pi^2_{12} = |\Phi^-\rangle_{12}\langle\Phi^-|,\
\Pi^3_{12} = |\Psi^+\rangle_{12}\langle\Psi^+|,\
\Pi^4_{12} = |\Psi^-\rangle_{12}\langle\Psi^-|,
\end{equation}
where
$$
|\Phi^{\pm}\rangle_{12} = \frac{1}{\sqrt{2}}(|00\rangle_{12} \pm |11\rangle_{12}),
$$
$$
|\Psi^{\pm}\rangle_{12} = \frac{1}{\sqrt{2}}(|01\rangle_{12} \pm |10\rangle_{12})
$$
are the Bell states.  If Alice's measurement has outcome $j$, she broadcasts \cite{Explanation1} her measurement result (two-bit) to Cindy and Bob via a classical channel.  The joint state of Cindy's system 3 and Bob's target system 4 conditioned on Alice's measurement result $j$ is given by
\begin{equation}
\rho^j_{34} = \frac{1}{p_j}{\rm tr}_{12}[(\Pi^j_{12} \otimes I_{34})(\pi_1 \otimes \chi^{GHZ}_{234})],
\end{equation}
where
\begin{equation}
p_j = {\rm tr}_{1234}[(\Pi^j_{12} \otimes I_{34})(\pi_1 \otimes \chi^{GHZ}_{234})].
\end{equation}
Substituting Eq.(1) to Eq.(3) into Eq.(5) yields $p_1 = p_2 = p_3 = p_4 = \frac{1}{4}$.  Next, Cindy performs a von Neumann measurement on system 3, described by operators $\Pi^k_3 \otimes I_4$, $I_4$ is the identity operator on subsystem 4, $k$ labels the outcome of the measurement,
$$
\Pi^1_3 = |\mu^+\rangle_3\langle\mu^+|,\ \Pi^2_3 = |\mu^-\rangle_3\langle\mu^-|,
$$
$$
|\mu^+\rangle_3 = \sin\nu|0\rangle_3 + \cos\nu|1\rangle_3,
$$
\begin{equation}
|\mu^-\rangle_3 = \cos\nu|0\rangle_3 - \sin\nu|1\rangle_3,
\end{equation}
$0 \leq \nu \leq \frac{\pi}{2}$.  If Cindy's measurement has outcome $k$, she communicates her measurement result (one-bit) to Bob via a classical channel.  The state of Bob's target system 4 conditioned on Cindy's measurement result $k$ is given by
\begin{equation}
\rho^{jk}_4 = \frac{1}{q_{jk}}{\rm tr_3}[(\Pi^k_3 \otimes I_4)\rho^j_{34}]
\end{equation}
where
\begin{equation}
q_{jk} = {\rm tr}_{34}[(\Pi^k_3 \otimes I_4)\rho^j_{34}].
\end{equation}
Substituting results from Eq.(4) and Eq.(6) into Eq.(8), we obtain
$$
q_{11} = q_{21} = q_{32} = q_{42} = \cos^2\frac{\theta}{2}\sin^2\nu + \sin^2\frac{\theta}{2}\cos^2\nu,
$$
\begin{equation}
q_{12} = q_{22} = q_{31} = q_{41} = \cos^2\frac{\theta}{2}\cos^2\nu + \sin^2\frac{\theta}{2}\sin^2\nu.
\end{equation}
For Bob to successfully complete the teleportation protocol, he performs a $j$- and $k$-dependent unitary operation $U^{jk}_4$ on system 4 (see Table I) such that
\begin{equation}
\tau^{jk}_4 = U^{jk}_4\rho^{jk}_4U^{jk\dagger}_4,
\end{equation}
where $U^{jk}$ could either be the identity matrix or one of the Pauli matrices:
$$
I = \left(\begin{array}{cc}
1 & 0 \\ 0 & 1
\end{array}\right),\ \sigma_x = \left(\begin{array}{cc}
0 & 1 \\ 1 & 0
\end{array}\right),\ \sigma_y = \left(\begin{array}{cc}
0 & -i \\ i & 0
\end{array}\right),\ \sigma_z = \left(\begin{array}{cc}
1 & 0 \\ 0 & -1
\end{array}\right).
$$
The success of the teleportation scheme can be measured by the fidelity \cite{Jozsa} between the input state $\pi_{in}$ and the output state $\tau^{jk}_{out}$, averaged over all possible Alice's and Cindy's measurement outcomes, $j$ and $k$ respectively, and over an isotropic distribution of input states $\pi_{in}$:
\begin{equation}
\langle F\rangle = 
\frac{1}{4\pi}\int^{\pi}_0\int^{2\pi}_0\sin\theta d\theta d\phi\
\sum^4_{j = 1}p_j\sum^2_{k = 1}q_{jk}F^{jk}
\end{equation}
where
\begin{equation}
F^{jk} \equiv {\rm tr}(\tau^{jk}_{out}\pi_{in}).
\end{equation}
It follows from Eq.(1) and results from Eq.(10) that
$$
F^{11} = F^{21} = F^{32} = F^{42} = \frac{(\cos^2\frac{\theta}{2}\sin\nu + \sin^2\frac{\theta}{2}\cos\nu)^2}{\cos^2\frac{\theta}{2}\sin^2\nu + \sin^2\frac{\theta}{2}\cos^2\nu},
$$
\begin{equation}
F^{12} = F^{22} = F^{31} = F^{41} = \frac{(\cos^2\frac{\theta}{2}\cos\nu + \sin^2\frac{\theta}{2}\sin\nu)^2}{\cos^2\frac{\theta}{2}\cos^2\nu + \sin^2\frac{\theta}{2}\sin^2\nu}.
\end{equation}
Substituting results from Eq.(5), Eq.(9) and Eq.(13) into Eq.(11) gives
\begin{equation}
\langle F\rangle = \frac{2}{3} + \frac{1}{3}\sin 2\nu.
\end{equation}
The average teleportation fidelity $\langle F\rangle$ is thus dependent on Cindy's von Neumann measurement on system 3, specified by $\nu$.  When $\nu = \frac{\pi}{4}$, $F^{jk} = 1$ for all $j$ and $k$, and we have $\langle F\rangle = 1$.

Now, we again consider the quantum teleportation protocol $P_0$ of A. Karlsson and M. Bourennane \cite{Karlsson}, but instead of Alice, Cindy and Bob sharing a three-particle entangled GHZ state, they share a three-particle entangled W state.  That is, the initial composite system 1234 is prepared in a state with density operator
$$
\sigma^{total}_{1234} = \pi_1 \otimes \chi^W_{234}
$$
where
\begin{equation}
\chi^W_{234} = |W\rangle_{234}\langle W|,\
|W\rangle_{234} = \frac{1}{\sqrt{3}}(|001\rangle_{234} + |010\rangle_{234} + |100\rangle_{234}).
\end{equation}
Consequently, the joint state of Cindy's system 3 and Bob's target system 4 conditioned on Alice's measurement result $j$ is given by
\begin{equation}
\tilde{\rho}^j_{34} = \frac{1}{\tilde{p}_j}{\rm tr}_{12}[(\Pi^j_{12} \otimes I_3)(\pi_1 \otimes \chi^W_{234})],
\end{equation}
where
\begin{equation}
\tilde{p}_j = {\rm tr}_{1234}[(\Pi^j_{12} \otimes I_3)(\pi_1 \otimes \chi^W_{234})].
\end{equation}
Substituting Eq.(1), Eq.(3) and Eq.(15) into Eq.(17) yields
$$
\tilde{p}_1 = \tilde{p}_2 = \frac{1}{6}(1 + \cos^2\frac{\theta}{2}),
$$
\begin{equation}
\tilde{p}_3 = \tilde{p}_4 = \frac{1}{6}(1 + \sin^2\frac{\theta}{2}).
\end{equation}
And, the state of Bob's target system 4 conditioned on Cindy's measurement result $k$ is given by
\begin{equation}
\tilde{\rho}^{jk}_4 = \frac{1}{\tilde{q}_{jk}}{\rm tr}_3[(\Pi^k_3 \otimes I_4)\tilde{\rho}^j_{34}]
\end{equation}
where
\begin{equation}
\tilde{q}_{jk} = {\rm tr}_{34}[(\Pi^k_3 \otimes I_4)\tilde{\rho}^j_{34}].
\end{equation}
Substituting Eq.(6) and results from Eq.(16) into Eq.(20) yields
$$
\tilde{q}_{11} = \frac{1 + \cos^2\frac{\theta}{2} - \sin^2\frac{\theta}{2}\cos2\nu + \sin\theta\cos\phi\sin2\nu}{2(1 + \cos^2\frac{\theta}{2})},
$$
$$
\tilde{q}_{12} = \frac{1 + \cos^2\frac{\theta}{2} + \sin^2\frac{\theta}{2}\cos2\nu - \sin\theta\cos\phi\sin2\nu}{2(1 + \cos^2\frac{\theta}{2})},
$$
$$
\tilde{q}_{21} = \frac{1 + \cos^2\frac{\theta}{2} - \sin^2\frac{\theta}{2}\cos2\nu - \sin\theta\cos\phi\sin2\nu}{2(1 + \cos^2\frac{\theta}{2})},
$$
$$
\tilde{q}_{22} = \frac{1 + \cos^2\frac{\theta}{2} + \sin^2\frac{\theta}{2}\cos2\nu + \sin\theta\cos\phi\sin2\nu}{2(1 + \cos^2\frac{\theta}{2})},
$$
$$
\tilde{q}_{31} = \frac{1 + \sin^2\frac{\theta}{2} - \cos^2\frac{\theta}{2}\cos2\nu + \sin\theta\cos\phi\sin2\nu}{2(1 + \sin^2\frac{\theta}{2})},
$$
$$
\tilde{q}_{32} = \frac{1 + \sin^2\frac{\theta}{2} + \cos^2\frac{\theta}{2}\cos2\nu - \sin\theta\cos\phi\sin2\nu}{2(1 + \sin^2\frac{\theta}{2})},
$$
$$
\tilde{q}_{41} = \frac{1 + \sin^2\frac{\theta}{2} - \cos^2\frac{\theta}{2}\cos2\nu - \sin\theta\cos\phi\sin2\nu}{2(1 + \sin^2\frac{\theta}{2})},
$$
\begin{equation}
\tilde{q}_{42} = \frac{1 + \sin^2\frac{\theta}{2} + \cos^2\frac{\theta}{2}\cos2\nu + \sin\theta\cos\phi\sin2\nu}{2(1 + \sin^2\frac{\theta}{2})}.
\end{equation}
Bob then performs a $j$- and $k$- dependent unitary operation $\tilde{U}^{jk}$ on system 4 (see Table II) such that
\begin{equation}
\tilde{\tau}^{jk}_4 = \tilde{U}^{jk}_4\tilde{\rho}^{jk}_4\tilde{U}^{jk\dagger}_4.
\end{equation}
The fidelity \cite{Jozsa} between the input state $\pi_{in}$ and the output state $\tilde{\tau}^{jk}_{out}$, averaged over all possible Alice's and Cindy's measurement outcomes, $j$ and $k$ respectively, and over an isotropic distribution of input states $\pi_{in}$ is therefore
\begin{equation}
\langle \tilde{F}\rangle = 
\frac{1}{4\pi}\int^{\pi}_0\int^{2\pi}_0\sin\theta d\theta d\phi\
\sum^4_{j = 1}\tilde{p}_j\sum^2_{k = 1}\tilde{q}_{jk}\tilde{F}^{jk}
\end{equation}
where
\begin{equation}
\tilde{F}^{jk} \equiv {\rm tr}(\tilde{\tau}^{jk}_{out}\pi_{in}).
\end{equation}
It follows from Eq.(1) and results from Eq.(22) that
$$
\tilde{F}^{11} = \frac{\frac{1}{2}\sin^2\theta\cos^2\nu + \sin\theta\cos\phi\sin2\nu + 2\sin^2\nu}{1 + \cos^2\frac{\theta}{2} - \sin^2\frac{\theta}{2}\cos2\nu + \sin\theta\cos\phi\sin2\nu},
$$
$$
\tilde{F}^{12} = \frac{2\cos^2\nu - \sin\theta\cos\phi\sin2\nu + \frac{1}{2}\sin^2\theta\sin^2\nu}{1 + \cos^2\frac{\theta}{2} + \sin^2\frac{\theta}{2}\cos2\nu - \sin\theta\cos\phi\sin2\nu},
$$
$$
\tilde{F}^{21} = \frac{\frac{1}{2}\sin^2\theta\cos^2\nu - \sin\theta\cos\phi\sin2\nu + 2\sin^2\nu}{1 + \cos^2\frac{\theta}{2} - \sin^2\frac{\theta}{2}\cos2\nu - \sin\theta\cos\phi\sin2\nu},
$$
$$
\tilde{F}^{22} = \frac{2\cos^2\nu + \sin\theta\cos\phi\sin2\nu + \frac{1}{2}\sin^2\theta\sin^2\nu}{1 + \cos^2\frac{\theta}{2} + \sin^2\frac{\theta}{2}\cos2\nu + \sin\theta\cos\phi\sin2\nu},
$$
$$
\tilde{F}^{31} = \frac{\frac{1}{2}\sin^2\theta\cos^2\nu + \sin\theta\cos\phi\sin2\nu + 2\sin^2\nu}{1 + \sin^2\frac{\theta}{2} - \cos^2\frac{\theta}{2}\cos2\nu + \sin\theta\cos\phi\sin2\nu},
$$
$$
\tilde{F}^{32} = \frac{2\cos^2\nu - \sin\theta\cos\phi\sin2\nu + \frac{1}{2}\sin^2\theta\sin^2\nu}{1 + \sin^2\frac{\theta}{2} + \cos^2\frac{\theta}{2}\cos2\nu - \sin\theta\cos\phi\sin2\nu},
$$
$$
\tilde{F}^{41} = \frac{\frac{1}{2}\sin^2\theta\cos^2\nu - \sin\theta\cos\phi\sin2\nu + 2\sin^2\nu}{1 + \sin^2\frac{\theta}{2} - \cos^2\frac{\theta}{2}\cos2\nu - \sin\theta\cos\phi\sin2\nu},
$$
\begin{equation}
\tilde{F}^{42} = \frac{2\cos^2\nu + \sin\theta\cos\phi\sin2\nu + \frac{1}{2}\sin^2\theta\sin^2\nu}{1 + \sin^2\frac{\theta}{2} + \cos^2\frac{\theta}{2}\cos2\nu + \sin\theta\cos\phi\sin2\nu}.
\end{equation}
Substituting Eq.(18), Eq.(21) and Eq.(25) into Eq.(23) gives
\begin{equation}
\langle \tilde{F}\rangle = \frac{1}{4\pi}\int^{\pi}_0\int^{2\pi}_0\sin\theta d\theta d\phi\ \frac{2}{3}(1 + \cos^2\frac{\theta}{2}\sin^2\frac{\theta}{2}) = \frac{7}{9} > \frac{2}{3},
\end{equation}
which is better than any classical communication protocol \cite{Bennett, Popescu, Horodecki}.  The average teleportation fidelity $\langle \tilde{F}\rangle$ is therefore, in contrast to Eq.(14), independent of Cindy's von Neumann measurement on system 3.  This is consistent with Table II, where we observe that Bob's necessary unitary operation $\tilde{U}^{jk}_4$ on system 4 depends only on Alice's measurement result $j$.  In fact, the same average teleportation fidelity can be obtained for both outputs at target systems 3 and 4, in a different teleportation protocol $P_1$, where Cindy plays the same receiver role as Bob:
\begin{equation}
\langle\tilde{F}_s\rangle = \frac{1}{4\pi}\int^{\pi}_0\int^{2\pi}_0\sin\theta d\theta d\phi\ \sum^4_{j = 1}\tilde{p}_j\tilde{F}^j_s
\end{equation}
where $s = 3$ or 4, and $\tilde{p}_j$ are as given in Eq.(18),
\begin{equation}
\tilde{F}^j_s \equiv {\rm tr}(\tilde{\tau}^{j(s)}_{out}\pi_{in}),\
\tilde{\tau}^{j(s)} = \left\{\begin{array}{c}
\tilde{U}^j({\rm tr}_4\tilde{\rho}^j_{34})\tilde{U}^{j\dagger}\ {\rm if}\ s = 3, \\
\tilde{U}^j({\rm tr}_3\tilde{\rho}^j_{34})\tilde{U}^{j\dagger}\ {\rm if}\ s = 4
\end{array}\right.
\end{equation}
with $\tilde{\rho}^j_{34}$ given by Eq.(16), and $\tilde{U}^j$ as in Table III.  Substituting Eq.(1) and results from Eq.(16) into Eq.(28) yields
$$
\tilde{F}^1_3 = \tilde{F}^1_4 = \tilde{F}^2_3 = \tilde{F}^2_4 = \frac{1 + \cos^2\frac{\theta}{2}\sin^2\frac{\theta}{2}}{1 + \cos^2\frac{\theta}{2}},
$$
\begin{equation}
\tilde{F}^3_3 = \tilde{F}^3_4 = \tilde{F}^4_3 = \tilde{F}^4_4 = \frac{1 + \cos^2\frac{\theta}{2}\sin^2\frac{\theta}{2}}{1 + \sin^2\frac{\theta}{2}}.
\end{equation}
It therefore follows from Eq.(18) and Eq.(29) that
\begin{equation}
\tilde{F}_3 = \tilde{F}_4 = \frac{7}{9}.
\end{equation}
This is in contrast to the average teleportation fidelity of $P_1$ using a three-particle entangled GHZ state instead \cite{Karlsson}, where
$$
\langle F_3\rangle = \langle F_4\rangle = \frac{2}{3}.
$$
However, we have to point out that generalizing the new protocol $P_1$ to one involving an $N$-particle entangled W state:
$$
\chi^W_{2\cdots N+1} = |W\rangle_{2\cdots N+1}\langle W|,
$$
\begin{equation}
|W\rangle_{2\cdots N+1} = \frac{1}{\sqrt{N}}(|0\cdots 01\rangle_{2\cdots N+1} + |0\cdots 010\rangle_{2\cdots N+1} + \cdots + |10\cdots 0\rangle_{2\cdots N+1})
\end{equation}
with
\begin{equation}
\tilde{\tau}^{j(s)} = \tilde{U}^j({\rm tr}_{3\cdots s-1\hat{s}s+1\cdots N + 1}\tilde{\rho}^j_{3\cdots N+1})\tilde{U}^{j\dagger},\ s = 3, \cdots, N + 1,
\end{equation}
where $\tilde{U}^j$ are as given in Table III,
$$
\tilde{\rho}^j_{3\cdots N+1} = \frac{1}{\tilde{p}_j}{\rm tr}_{12}[(\Pi^j_{12} \otimes I_{3\cdots N+1})(\pi_1 \otimes \chi^W_{23\cdots N+1})],
$$
$$
\tilde{p}_1 = \tilde{p}_2 = \frac{1}{2N}[1 + (N - 2)\cos^2\frac{\theta}{2}],
$$
\begin{equation}
\tilde{p}_3 = \tilde{p}_4 = \frac{1}{2N}[1 + (N - 2)\sin^2\frac{\theta}{2}],
\end{equation}
which give via Eq.(28),
$$
\tilde{F}^1_s = \tilde{F}^2_s = \frac{1 + (N - 2)\cos^2\frac{\theta}{2}\sin^2\frac{\theta}{2}}{1 + (N - 2)\cos^2\frac{\theta}{2}},
$$
\begin{equation}
\tilde{F}^3_s = \tilde{F}^4_s = \frac{1 + (N - 2)\cos^2\frac{\theta}{2}\sin^2\frac{\theta}{2}}{1 + (N - 2)\sin^2\frac{\theta}{2}}
\end{equation}
and therefore,
\begin{equation}
\langle\tilde{F}_s\rangle = \frac{N + 4}{3N}.
\end{equation}
It is clear from Eq.(35) that $\langle\tilde{F}_s\rangle \leq \frac{2}{3}$ for $N \geq 4$.  This is consistent with the picture that in such a scheme as $P_1$, the information encoded in $\pi_{in}$ is evenly distributed among the receivers, and is ``spread even thinner'' with an increasing number of such receivers.  Naturally, we would expect the average teleportation fidelity to deteriorate, in agreement with the no-cloning theorem \cite{Zurek}.  So, if one wishes to have an output state with an average fidelity greater than $\frac{2}{3}$, one would have to modify $P_1$ such that it involves less number of receivers.  Therefore, measurement of the sort in $P_0$ is necessary \cite{Illustration}.  As a matter of fact, Cindy's measurement in $P_0$ does allow Bob to obtain a perfect copy of the input state, albeit only probabilistically.  For instance, when Alice's measurement result is $j = 1$,  Cindy could perform a von Neumann measurement on system 3 with $\nu = \frac{\pi}{2}$, then with probability, Eq.(21),
$$
\tilde{q}_{11} = \frac{1}{1 + \cos^2\frac{\theta}{2}}
$$
Bob could receive a perfect copy with $\tilde{F}^{11} = 1$.

Finally, we study the effects on the average fidelity of the Karlsson-Bourennane teleportation protocol $P_0$ \cite{Karlsson} due to the presence of white noise.  We want to compare the effect on the three-particle GHZ state with that on the three-particle W state.  To this end, we consider
\begin{equation}
\hat{\chi}^{GHZ}_{234} = w\chi^{GHZ}_{234} + (1 - w)\frac{1}{8}I_{234},
\end{equation}
and
\begin{equation}
\hat{\chi}^W_{234} = w\chi^W_{234} + (1 - w)\frac{1}{8}I_{234},
\end{equation}
where $0 \leq w \leq 1$ is called visibility in Ref.\cite{Dagomir}.  It defines to what extent the quantum processes associated with $\chi^{GHZ}_{234}$ $(\chi^W_{234})$ are visible in those given by $\hat{\chi}^{GHZ}_{234}$ $(\hat{\chi}^W_{234})$.  If $w = 0$, no trace is left of these $\chi^{GHZ}_{234}$ $(\chi^W_{234})$ generated processes, and if $w = 1$, we have the full visibility, not affected by any noise.  Relpacing $\chi^{GHZ}_{234}$ in Eq.(4) and $\chi^W_{234}$ in Eq.(16) with Eq.(36) and Eq.(37) respectively, and going through essentially the same calculations, we obtain for the three-particle GHZ state,
\begin{equation}
\langle F\rangle = \frac{1}{2} + \frac{1 + 2\sin2\nu}{6} - \frac{1 + 2\sin2\nu}{6}(1 - w),
\end{equation}
and for the three-particle W state,
\begin{equation}
\langle\tilde{F}\rangle = \frac{7}{9} - \frac{5}{18}(1 - w).
\end{equation}
Once again, $\langle\tilde{F}\rangle$ is independent of Cindy's measurement on system 3, and it decreases at a constant rate with resepct to $(1 - w)$.  More interestingly, the rate of decrease of $\langle F\rangle$ with respect to $(1 - w)$ depends on Cindy's measurement on system 3, specified by $\nu$.  $\langle F\rangle$ decreases at a higher rate when Cindy's measurement yields a higher $\langle F\rangle$.

In conclusion, we recast the teleportation scheme $P_0$ of A. Karlsson and M. Bourennane \cite{Karlsson} in the language of density operators and quantum operations.  This allows us to investigate the consequences of replacing the three-particle GHZ state in their original scheme with a three-particle W state.  We compare and contrast our results with theirs.  We are also able to study the effects of white noise on the average teleportation fidelity.  However, it is not clear if $P_0$ (or $P_1$) is optimal, i.e., if there exist more general completely positive maps rather than the unitary operators $\tilde{U}^{jk}$ (or $\tilde{U}^j$), which could yield a higher average teleportation fidelity.  It would also be interesting to see how the average fidelities would change when the GHZ and W states are being subjected to other types of noise.

The author thanks Yuri Suhov and Andrew Skeen for useful discussions.  This publication is an output from project activity funded by The Cambridge MIT Institute Limited (``CMI'').  CMI is funded in part by the United Kingdom Government.  The activity was carried out for CMI by the University of Cambridge and Massachusetts Institute of Technology.  CMI can accept no responsibility for any information provided or views expressed.

\newpage

\begin{table}
\begin{ruledtabular}
\begin{tabular}{ccc}
Alice's measurement result $j$ & Cindy's measurement result $k$ & Bob's unitary operation $U^{jk}$\\
1 & 1 & $I$ \\
1 & 2 & $\sigma_z$ \\
2 & 1 & $\sigma_z$ \\
2 & 2 & $I$ \\
3 & 1 & $\sigma_x$ \\
3 & 2 & $\sigma_y$ \\
4 & 1 & $\sigma_y$ \\
4 & 2 & $\sigma_x$
\end{tabular}
\end{ruledtabular}
\caption{\label{I}}
Bob's unitary operations conditioned on both Alice's and Cindy's measurement results, when Alice, Cindy and Bob share a three-particle entangled GHZ state.
\end{table}

\begin{table}
\begin{ruledtabular}
\begin{tabular}{ccc}
Alice's measurement result $j$ & Cindy's measurement result $k$ & Bob's unitary operation $\tilde{U}^{jk}$\\
1 & 1 & $\sigma_x$ \\
1 & 2 & $\sigma_x$ \\
2 & 1 & $\sigma_y$ \\
2 & 2 & $\sigma_y$ \\
3 & 1 & $I$ \\
3 & 2 & $I$ \\
4 & 1 & $\sigma_z$ \\
4 & 2 & $\sigma_z$
\end{tabular}
\end{ruledtabular}
\caption{\label{II}}
Bob's unitary operations conditioned on both Alice's and Cindy's measurement results, when Alice, Cindy and Bob share a three-particle entangled W state.
\end{table}

\newpage

\begin{table}
\begin{ruledtabular}
\begin{tabular}{ccc}
Alice's measurement result $j$ & Bob's unitary operation $\tilde{U}^{j}$ & Cindy's unitary operation $\tilde{U}^{j}$ \\
1 & $\sigma_x$ & $\sigma_x$\\
2 & $\sigma_y$ & $\sigma_y$\\
3 & $I$ & $I$\\
4 & $\sigma_z$ & $\sigma_z$
\end{tabular}
\end{ruledtabular}
\caption{\label{III}}
Bob's and Cindy's unitary operations conditioned only on  Alice's measurement result, when Alice, Cindy and Bob share a three-particle entangled W state.  Cindy does not perform any von Neumann measurement on system 3.
\end{table}


\begin{thebibliography}{99}
\bibitem{Explanation1} In $P_0$, Cindy needs to know Alice's measurement outcomes so that if $j = 3$ or 4 she could rotate, after her measurement on system 3, the state of system 3 with $\sigma_x$ such that the average fidelity between the state of system 3 and the input state is $\frac{2}{3} - \frac{1}{6}\sin^22\nu$.  Since this is never greater than $\frac{2}{3}$, we do not discuss it in the main text.
\bibitem{Illustration} It would be interesting to see if there is a general strategy of measurements which would yield any number of outputs with desired average fidelities.
\end{thebibliography}
\end{document}